\documentclass [12pt]{article}
\usepackage{graphics,epsfig}
\title{Radiation stability of biocompatibile magnetic fluid}
\author{Nat\'alia Toma\v{s}ovi\v{c}ov\'a$^1$, Ivan Haysak$^2$, Martina Konerack\'a$^1$, \\
Jozef Kov\'a\v{c}$^1$, Milan Timko$^1$, Vlasta Z\'avi\v{s}ov\'a$^1$, Alexander Okunev$^2$, \\
Alexander Parlag$^2$, Alexey Fradkin$^2$, Peter Kop\v{c}ansk\'y$^1$ \\
$^1$Institute of Experimental Physics, Slovak Academy of Sciences, \\ 
Watsonov\'a 47, 04001 Ko\v{s}ice, Slovakia \\
$^2$Uzhgorod National University, Kapitulna 9A, 880 00 Uzhgorod, Ukraine}\date{}
\begin{document}
\baselineskip=20pt
\maketitle

\begin{abstract}
The radiation stability of biocompatibile magnetic fluid used in
nanomedicine after electron irradiation was studied. Two types of
the water-based magnetic fluids were prepared. The first one was
based on the magnetite nanoparticles stabilized by one  surfactant
natrium oleate. The second one was biocompatibile magnetic fluid
stabilized with two surfactants, natrium oleate as a first
surfactant and Poly(ethylene glycol) (PEG) as a second surfactant.
The magnetization measurements showed that electron irradiation up
to 1000Gy caused 50\% reduction of saturation magnetization in the
case of the first sample with only one surfactant  while in the case
of the second biocompatibile magnetic fluid,  only 25\% reduction of
saturation magnetization was observed. In the first magnetic fluid
the radiation  caused the  higher sedimentation   of the magnetic
particles than in the second case, when magnetic particles are
covered also with PEG. The obtained results show that PEG behave as
a protective element.
\end{abstract}

\newpage

\section{Introduction}

Nanotechnology is beginning to allow scientist, engineers and
physicians to work at the cellular and molecular levels to produce
major advances in the life sciences and healthcare. Ferrofluids or
magnetic fluids (dispersions of magnetic nanoparticles in liquids)
belong to complex systems widely studied in modern nanoscience.
Magnetic nanoparticles show remarkable phenomena such as
superparamagnetism, high field irreversibility, high saturation
field that make them very attractive for applications in biomedicine
for example in drug targeting delivery, magnetic hyperthermia,
arrangement of biological assemblies, contrast agents in Magnetic
Resonance Imaging, biomagnetic separation ~\cite{P}. Biomedical
applications require the magnetic particles to be stable in water at
neutral pH and physiological salinity. The colloidal stability of
magnetic fluids ~\cite{APH,ESK} will depend on the dimensions of
particles, which should be sufficiently small to avoid of
aggregation and on the surfactant commonly a monolayer of oleic acid
(steric repulsion), or the particles are prevented from sticking to
each other by electrostatic bilayer (electrostatic repulsion)
~\cite{MLE,MA1,DB,MA2}. Moreover, for $in$ $vivo$ applications the
magnetic particles must be coated with biocompatibile polymer
~\cite{SIP,CH}.
 Magnetic  hyperthermia is one of the widely studied application
 of magnetic fluids in medicine ~\cite{IA,AJ,MJ1,MJ2}.
 In work ~\cite{MJ2}, the combined thermotherapy and radiation
 with 20~Gy was shown to be significantly more effective than radiation with  20~Gy alone.
 Up to now there is very poor information about influence
of radiation on the physical properties of the magnetic fluids. To
our knowledge there is one  publication ~\cite{PK} where influence
of gamma radiation on magnetic properties of magnetic particles in
the kerosene-based magnetic fluids. From the point of basic and
applied research it is very important to study the physical
properties and colloidal stability of magnetic fluids after
radiation  that are used for bioaplications.  The aim of this paper
was to study the stability of the magnetic properties of two types
of water-based magnetic fluids suitable for bioaplications.

\section{Experiment}

The magnetic Fe$_3$O$_4$ particles were synthesized by a chemical
co-precipitation procedure. Ferrous chloride heptahydrate
(FeSO$_4$·7H$_2$O) and ferric chloride hexahydrate
(FeCl$_3$·6H$_2$O) with molar ratio of 1:2 were dissolved in
deionized water under vigorous stirring. Then 25~\% w/v ammonium
hydroxide (NH$_4$OH) was added as a base source at room temperature
until pH value of 11 was reached. The sediment obtained was washed
for five times with deionized water to remove impurities. The
stabilization of the magnetite precipitate was achieved by adding
natrium oleate (C$_{17}$H$_{33}$COONa), ratio 1.1 g to 1.5 g of
Fe$_3$O$_4$ during stirring and heating until the boiling point was
reached. Then centrifugation at 9000 rpm during 30~min followed. The
system at this stage is investigated below as initial magnetic
fluid. In a second magnetic fluid the poly(ethylene glycol) (PEG
purchased from Sigma company) with chemical
 formula C$_{18}$H$_{33}$NaO$_2$  was used as
a second surfactant with aim to increase the stability and improve
the biocompatibility  of the magnetic particles. Adsorption of PEG
was carried out by adding PEG (Mw = 1000) to the magnetic fluid in
the form of 10~\% w/v
 water solution, while it was stirred and heated up to  50 $^\circ$C. The mixture was stirred for one
hour and then left to cool down to a room temperature.  The added
amount was 0.25~g of PEG per 1~g of Fe$_3$O$_4$. The prepared
magnetic fluid contain  105~mg Fe$_3$O$_4$ /ml and the pH was 10.24.

\begin{figure}
\begin{center}
\includegraphics[width=20pc]{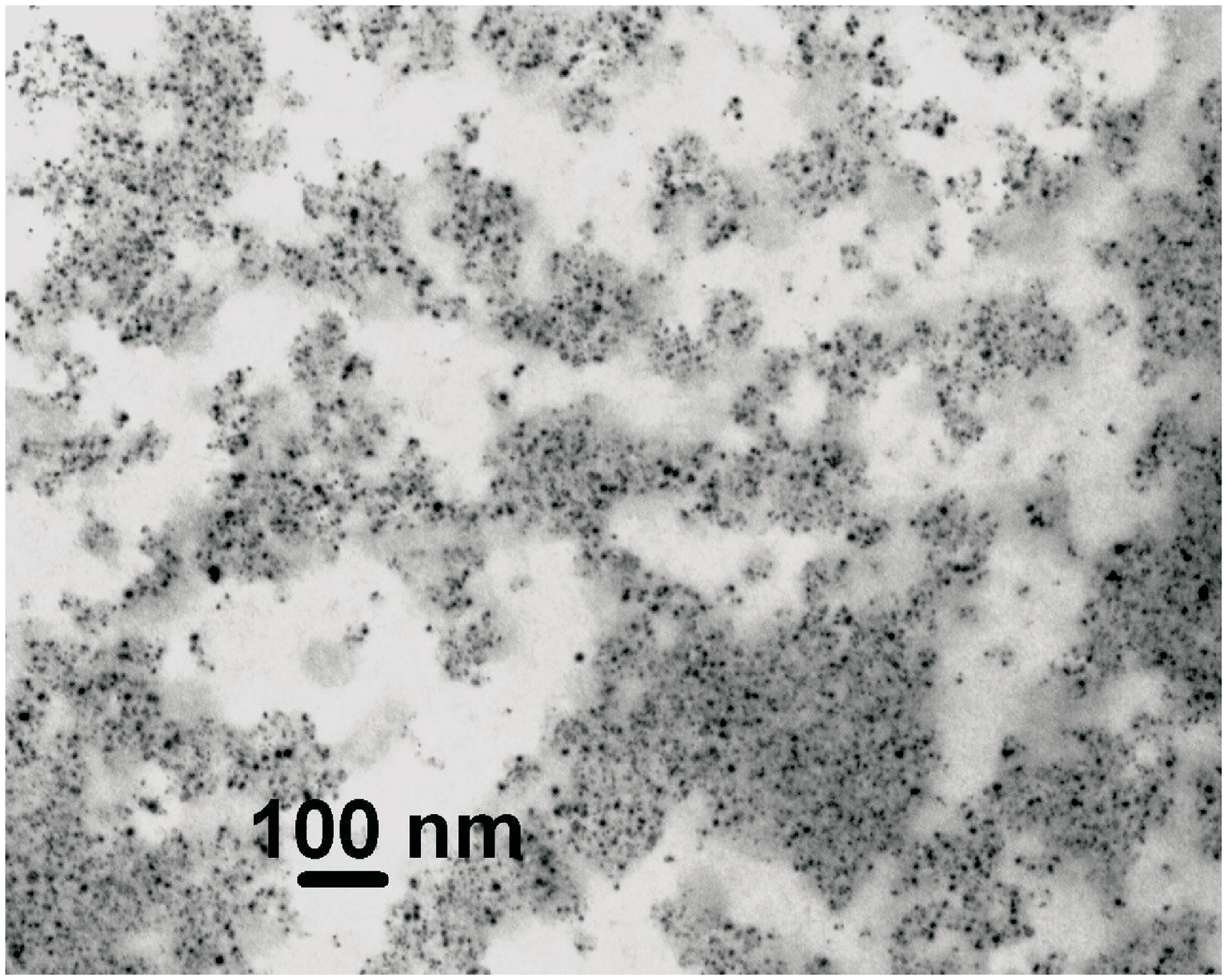}
\caption{TEM image of magnetite nanoparticles.} 
\label{TEM}
\end{center}
\end{figure}

\begin{figure}
\begin{center}
\includegraphics[width=20pc]{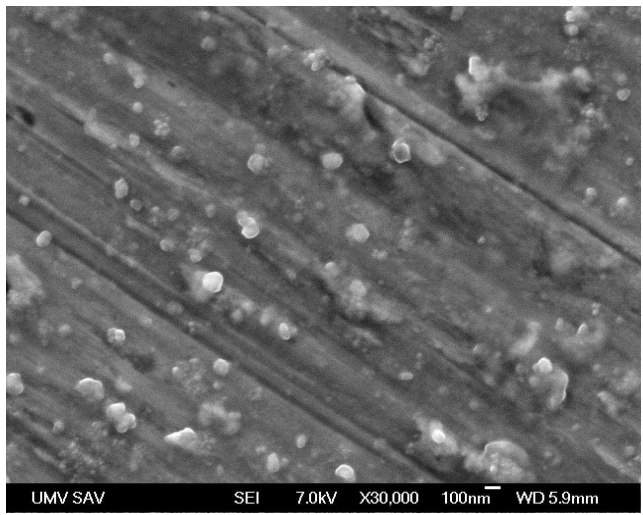}
\caption{\label{SEM} SEM image of magnetic nanoparticles coated with
natrium oleate and PEG.}
\end{center}
\end{figure}

With the aim to confirm the immobilization of natrium oleate and PEG
to the magnetic particles,  Fourier transform infrared (FTIR)
spectra of the pure natrium oleate and PEG and   after adsorption on
the magnetite  were measured by FTIR spectrometer FTLA2000
instrument (ABB, resolution 4~cm$^{-1}$) by the KBr pellet method.
In this method, the solid sample is finely pulverized with pure and
dry KBr, the mixture is pressed in a hydraulic press to form a
transparent pellet, and the spectrum of the pellet is measured.
Morphology and size distribution of prepared samples were obtained
from transmission electron microscopy  and Scanning electron
microscopy (SEM). TEM showed nearly spherical shape of the magnetite
core of the prepared magnetic fluids with average diameter 5 nm
(Fig.~\ref{TEM}). Fig.~\ref{SEM} shows the SEM image of the magnetic
particles coated with natrium oleate and PEG.

The irradiation of samples was conducted by electrons on an
accelerator "Microtron M-10" of Uzhgorod National University. The
beam of electrons with energy 8.6 MeV was deflected in atmosphere
through a thin titanic window. The samples in a glass container were
exposed in the distance of 40 cm from the output window of the
microtron. An intensity of the beam on this distance was set within
the limits of $\Phi$=(1-8) $\times$ 10$^9$ electrons/cm$^2$ and was
controlled by a thin-walled ionization chamber. The ionization
chamber was calibrated by a Faraday cup the entrance window of which
was disposed in the place of exposed patterns. This ionization
chamber was used for measuring of dose in the process of irradiation
too. For this purpose the output of the chamber was connected to an
integrator of current. The value of absorbed dose was estimated by
multiplying of electron fluence to a factor 3.3 $\times$ 10$^{-9}$
Gy cm$^2$. The intensity of the electron beam during the process of
irradiation remained stable within the limits of  $\pm$ 5\%.

The magnetization curves of the samples before irradiation and after
irradiation with different doses were measured  by SQUID
magnetometer (Quantum Design MPMS 5XL) twice.  The second
measurement was done one month after first measurement. The upper
part of the samples, kept in screw capped vials, was withdrawn for
measurements. The weight of measured samples varied from the
interval 20mg - 30mg. The error of  weight was 4\% that influence
values of magnetization measurements.

The FTIR spectra of the  samples before and after irradiation were
measured by FTIR spectrometer FTLA2000 instrument (ABB, resolution
4~cm$^{-1}$) by Attenuated Total Reflectance measurements with
diamond window. For infrared measurements the samples were first
mixed and than dried.

\section{Results and discussion}
The prepared samples, water-based magnetic fluid containing magnetic
particles coated with natrium oleate and water-based magnetic fluid
containing magnetic particles coated with natrium oleate as a first
surfactant and PEG as a second surfactant were irradiated with
different fluences 1.65$\times$10$^{10}$, 6.6$\times$10$^{10}$,
3.3$\times$10$^{11}$, 1.65$\times$10$^{12}$, 3.3$\times$10$^{12}$
electrons/cm$^2$ that corresponds doses 5Gy, 20Gy, 100Gy, 500Gy and
1000Gy, respectively.

\begin{figure}%
\begin{center}
\includegraphics*[width=30pc]{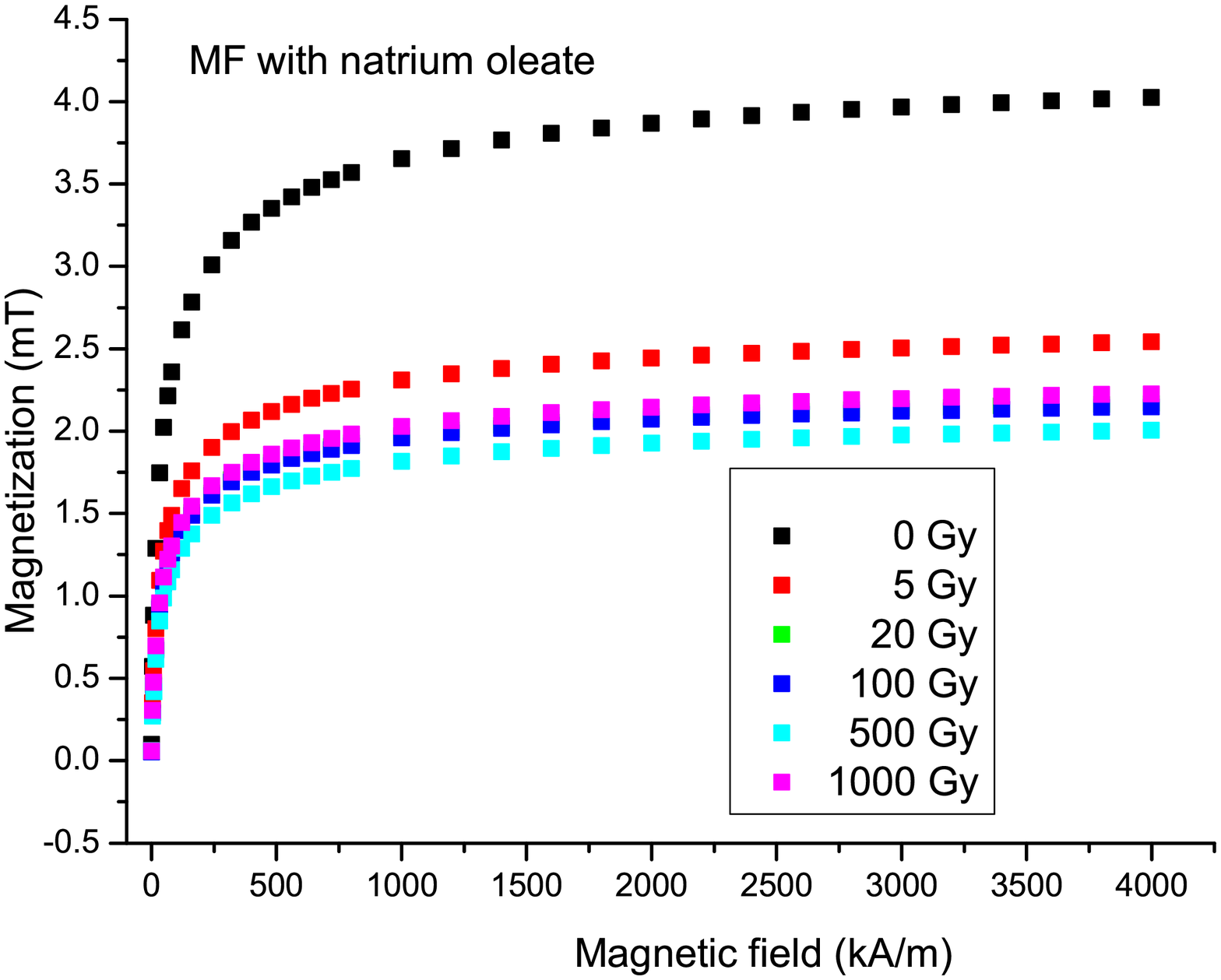}
\caption{Magnetization curves of the water-based magnetic
fluid containing the magnetic particles coated with natrium oleate
before and after irradiation with different doses.} 
\label{b-MF-n}
\end{center}
\end{figure}

\begin{figure}%
\includegraphics*[width=30pc]{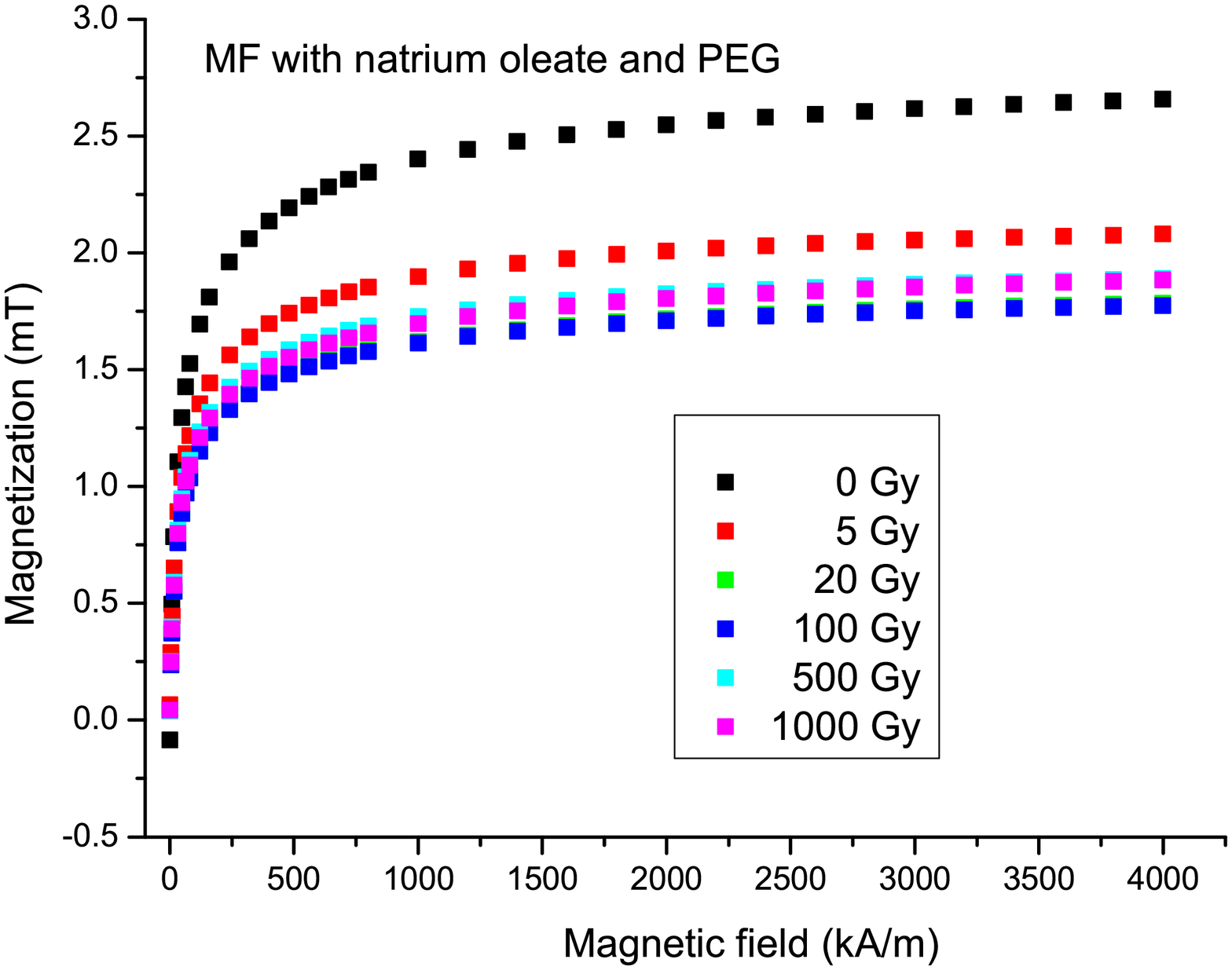}
\caption{Magnetization curves of the water-based magnetic
fluid containing the magnetic particles coated with natrium oleate
and PEG before and after irradiation with different doses.}
\label{a-MF-n}
\end{figure}

All studied  samples were placed in the screw capped vials. One week
after irradiation the upper parts of magnetic fluids were withdrawn
for magnetisation measurements. The magnetization curves  of the
samples before irradiation and after irradiation with different
doses of the magnetic fluid containing magnetic nanoparticles coated
with one surfactant (natrium oleate) and magnetic fluid containing
magnetic particles coated with two surfactants (natrium oleate and
PEG)are shown in Fig.~\ref{b-MF-n} and Fig.~\ref{a-MF-n},
respectively. The obtained results shoved significant reduction of
magnetization after irradiation of 5Gy. The next increase of the
dose influence the magnetization only slightly. The similar
behaviour was observed for both magnetic fluids.

\begin{figure}%
\begin{center}
\includegraphics*[width=30pc]{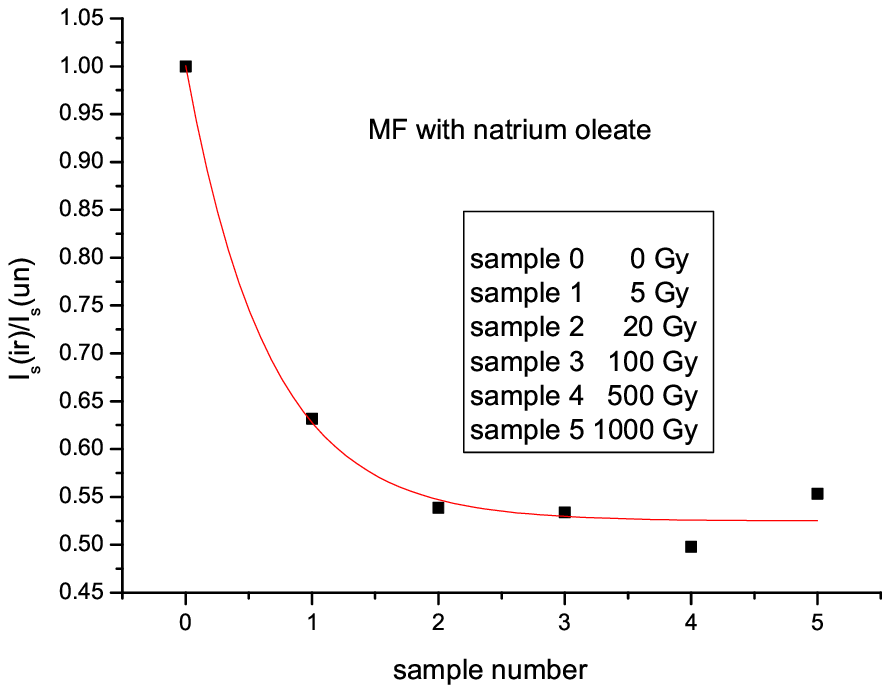}
\caption{Dependence of reduced saturation magnetization on
irradiation doses of water-based  magnetic fluid containing magnetic
particles coated with natrium oleate. The solid line is for eyes.}
\label{B}
\end{center}
\end{figure}

\begin{figure}%
\begin{center}
\includegraphics*[width=30pc]{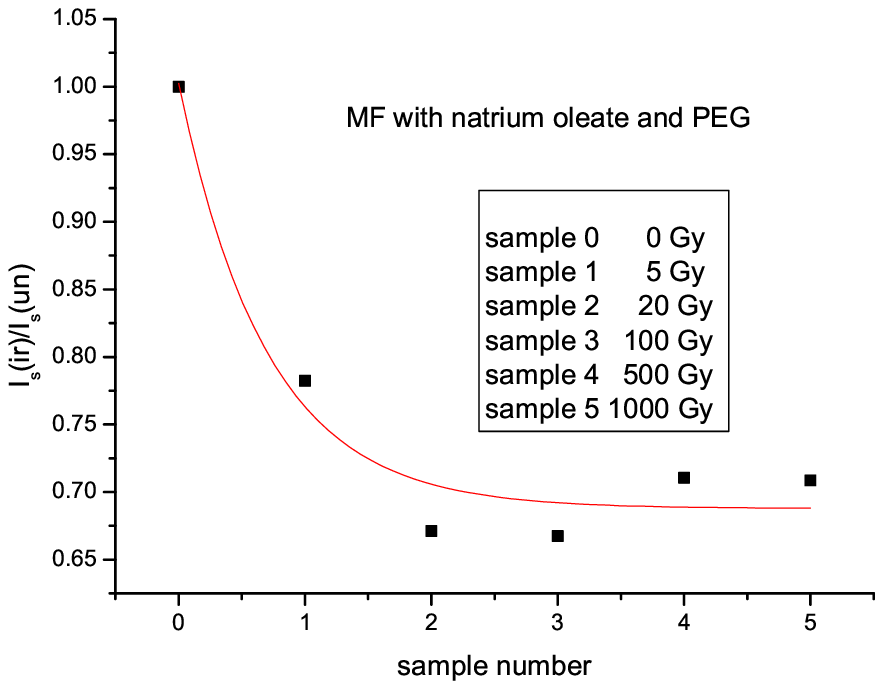}
\caption{Dependence of reduced saturation magnetization on
irradiation doses of water-based  magnetic fluid containing magnetic
particles coated with natrium oleate and PEG. The solid line is for
eyes.}
\label{A}
\end{center}
\end{figure}

Fig.~\ref{B} and Fig.~\ref{A} show the reduction of the saturated
magnetization for the irradiated samples. From this figures is
clearly seen that the electron
  irradiation up to
20 Gy caused 50\% reduction of the saturation magnetization in the
case of the  sample with only one surfactant  while in the case of
the second biocompatibile magnetic fluid coated also with PEG,  only
25\% reduction of the saturation magnetization was observed. The
obtained results show that in the first magnetic fluid the
irradiation caused the higher sedimentation of the magnetic
particles than in the second one, when magnetic particles are
covered also with PEG. From these figures is also seen that there is
no additional reduction in the saturated magnetization after
irradiation with higher doses up to 1000Gy.

The second measurement of the magnetisation was done one month after
the first measurement. The samples for the measurement were prepared
the same way as for the first measurement.  In Tab.~\ref{tab1} and
Tab.~\ref{tab2} are summarized  the saturated magnetizations
obtained  from the first and from the second  magnetisation
measurements of the sample with one surfactant and the sample with
two surfactants, respectively. The obtained results show no change
in the saturation magnetization after one month. These results show
that after sedimentation process that occur after irradiation, there
is no additional sedimentation and all samples are time stable.

\begin{table}
\caption{\label{tab1}The saturated magnetization (in mT) of the
sample with one surfactant before and after irradiation obtained
from the first measurements and from the measurements after one
month.} \vspace{0.2cm}
\begin{tabular}{lcccccc}
\hline
dose &  0Gy & 5Gy & 20Gy &  100Gy & 500Gy & 1000Gy \\
\hline
 1$^{st}$  & 4.02 & 2.54 & 2.16 & 2.14 & 2.08 & 2.22 \\
2$^{nd}$ & 4.01 & 2.55 & 2.18 & 2.12 & 2.10 & 2.10 \\
\hline
\end{tabular}
\end{table}

\begin{table}
\caption{\label{tab2}The saturated magnetization (in mT) of the
sample with two surfactants before and after irradiation from the
first measurements and form the measurements after one month.}
\vspace{0.2cm}
\begin{tabular}{lcccccc}
\hline
dose &  0Gy & 5Gy & 20Gy &  100Gy & 500Gy & 1000Gy \\
\hline
 1$^{st}$  & 2.65 & 2.07 & 1.78 & 1.77 & 1.88 & 1.88 \\
2$^{nd}$ & 2.65 & 2.10 & 1.85 & 1.75 & 1.74 & 1.82 \\
\hline
\end{tabular}
\end{table}

As possible mechanisms of magnetism degradation after irradiation
can be considered nuclear reactions, ionization processes and
degradation of surfactant molecules. However, there are no nuclear
reaction at the used electron energy 8MeV.  So the only possible
processes at the used electron energy are ionization and molecules
destroying that could lead to the aggregation of the particles. On
the other hand, the decrease of magnetization has saturating
behaviour and after sedimentation that occurs after irradiation
there is no additional sedimentation and samples are time stable.

\begin{figure}%
\begin{center}
\includegraphics*[width=30pc]{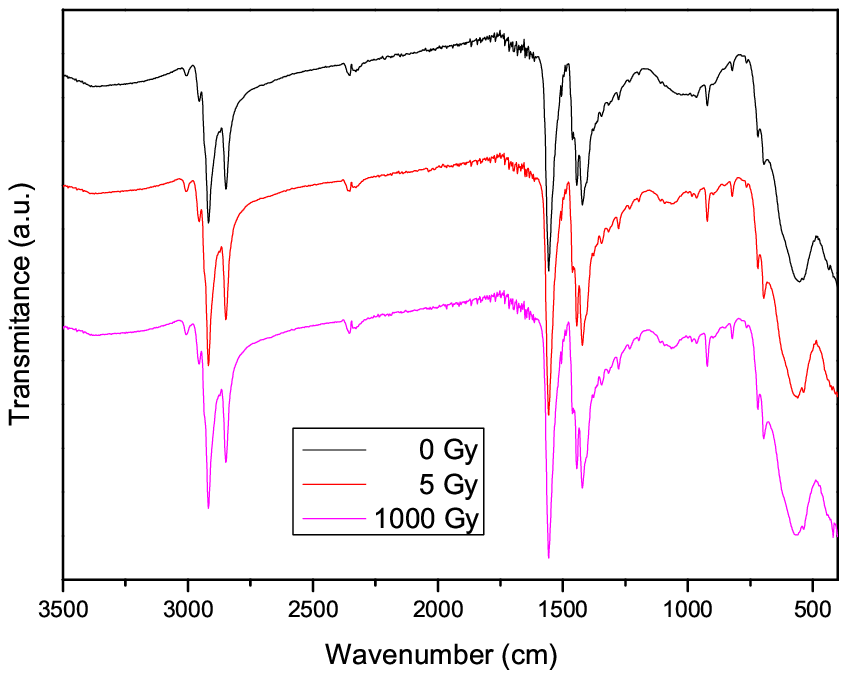}
\caption{FTIR spectra of magnetite coated with natrium oleate before irradiation and after irradiation with 5Gy and
  1000Gy. The spectra are shifted vertically for clarity.} 
\label{IR-B}
\end{center}
\end{figure}

\begin{figure}%
\begin{center}
\includegraphics*[width=30pc]{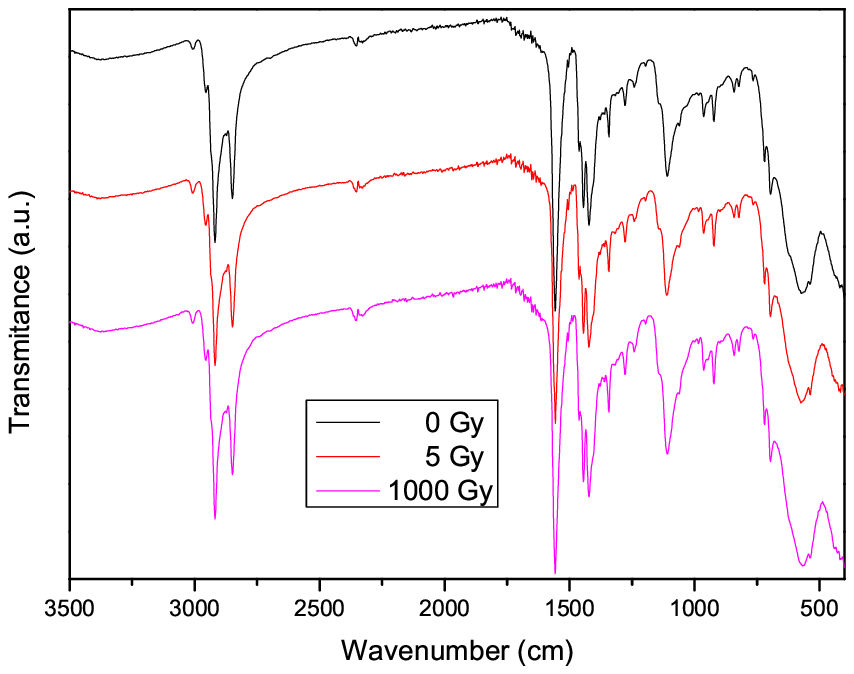}
\caption{FTIR spectra of magnetite coated with natrium oleate and PEG before irradiation and after irradiation with 5Gy and
  1000Gy. The spectra are shifted vertically for clarity.} 
\label{IR-A}
\end{center}
\end{figure}

Fig.~\ref{IR-B} and Fig.~\ref{IR-A} show the infrared spectra of
magnetic particles coated only with natrium oleate and magnetic
particles coated with natrium oleate and PEG, respectively. The
spectra are shifted vertically for clarity. The all absorption bands
corresponding the natrium oleate or PEG as well as  the band
observed at 584~cm$^{-1}$, which corresponds to the magnetite,
before irradiation and after irradiation are identical. These
results clearly indicate that the molecules of the surfactants as
well as the magnetite particles are stable under irradiation and no
destroying processes occur due to the irradiation. It seems that
ionization can be mainly responsible for aggregation of the
particles.

\section{Conclusion}
In summary, the magnetization measurements showed that electron
irradiation up to 1000Gy caused 50\%
  reduction of saturation magnetization in the case of the  sample with only one surfactant. In the case of
  the second biocompatibile magnetic fluid,  only 25 \% reduction of the saturation magnetization was observed.
  The obtained results showed that the radiation  causes aggregation of the particles and consequently their
   sedimentation.
The  infrared spectra confirmed  that the process that causes the
sedimentation after irradiation is ionization.
   However, in the  case  when magnetic particles were covered also with PEG, the obtained
  results showed that PEG behave as a protective element. Moreover,
  after first sedimentation process that occurs after irradiation, there is no further sedimentation and
 the sample with one surfactant as well as the sample with two surfactants  are time stable.

\section*{Acknowledgments}
This work was supported by the  Ministry of Education and Science of
Ukraine in the framework of project DB-715,
  by the Slovak Academy of Sciences, in the framework of CEX-NANOFLUID,
projects VEGA 0077 and 0311, APVV 0509-07 and Ministry of Education
Agency for Structural Fonds of EU in frame of project 6220120021 and
26220220005.

\newpage

\end{document}